\documentclass[11pt,a4paper]{article}
\usepackage{amsmath}
\usepackage[T1]{fontenc}
\usepackage[utf8]{inputenc}
\usepackage{authblk}
\usepackage{graphicx}
\usepackage{subfigure}
\usepackage{color}
\usepackage[numbers,sort&compress,square]{natbib}

\title{%
Elastic properties of mono-- and polydisperse two--dimensional crystals of
hard--core repulsive Yukawa particles}

\author[1]{Jakub W. Narojczyk}
\author[1]{Pawe{\l} M. Pig{\l}owski}
\author[2]{Krzysztof W. Wojciechowski}
\author[1]{Konstantin~V. Tretiakov\thanks{kvt@ifmpan.poznan.pl}}
\affil[1]{Institute of Molecular Physics, Polish Academy of
Sciences, Smoluchowskiego 17, 60-179 Pozna\'n, Poland}
\affil[2]{President St. Wojciechowski PWSZ in Kalisz,
Nowy Swiat 4, 62-800 Kalisz, Poland}

\begin{document}

\maketitle

\begin{abstract}%
Monte Carlo simulations of mono-- and polydisperse two--dimensional crystals are
reported. The particles in the studied system, interacting through hard--core
repulsive Yukawa potential, form a solid phase of hexagonal lattice. The elastic
properties of crystalline Yukawa systems are determined in the $NpT$ ensemble
with variable shape of the periodic box. Effects of the Debye screening length
($\kappa^{-1}$), contact value of the potential ($\epsilon$), and the size
polydispersity of particles on elastic properties of the system are studied.
The simulations show that the polydispersity of particles strongly influences 
the elastic properties of the studied system, especially on the shear modulus. 
It is also found that the elastic moduli increase with density and their growth 
rate depends on the screening length. Shorter screening length leads to faster
increase of elastic moduli with density and decrease of the Poisson's ratio.
In contrast to its three-dimensional version, the studied system is non-auxetic, 
i.e. shows positive Poisson's ratio.
\end{abstract}

\section{Introduction}
Colloidal suspensions are systems consisting of small particles of sizes from
tens to hundreds nanometers dispersed in a solvent~\cite{GasRus98PT}. The 
particles can form colloidal crystals not only in three dimensions
(3D)~\cite{RusETAL91BOOK} but also in two dimensions (2D) where they create a
hexagonal lattice~\cite{Pie80PRL}. When particles are endowed with
electrical charge and placed in a dispersing medium, they exhibit exponential
decay of electric potential due to screening effect from ions in the
solvent~\cite{AueFre2002JPCM}. This decay is described by the \emph{screening
length} $\kappa^{-1}$ (known also as the \emph{Debye length}) which depends on
temperature, permittivity, and concentration of ions in the solvent.
Interactions between particles in such systems are decribed by Hard--Core 
Repulsive Yukawa Potential (HCRYP)~\cite{AzhBauRyc2000JCP,HynDij2003JPCM}, 
both in two~\cite{AsgDavTan2001PRE} and three~\cite{AzhBauRyc2000JCP,%
HynDij03PRE} dimensions.
The potential consists of two parts: the hard core
(reflecting the influence of finite sizes of particles) and the repulsive
part describing spherically symmetric Coulomb screening effect. HCRYP systems
can behave (depending on $\kappa^{-1}$) as hard particle systems (perfect
screening) or as Wigner crystal (no screening)~\cite{AssMesLow2008JCP}.

More than ten years ago the liquid--solid phase diagrams for two-dimensional 
(2D)~\cite{AsgDavTan2001PRE} and three-dimensional (3D)~\cite{AueFre2002JPCM,%
AzhBauRyc2000JCP,DavKohMoh2000PRE,DixZuk2003JPCM} HCRYP systems have been 
studied. The research showed that in 2D HCRYP particles can crystallize into a 
hexagonal lattice, whereas the in 3D they form f.c.c. or b.c.c. structures 
(depending on external conditions). Various
thermodynamic properties and physical phenomena have been studied for the Yukawa 
potential, both in 2D~\cite{AsgDavTan2001PRE,AssMesLow2008JCP,GlaLow2012JPCM,%
HarKalDon2005PRE,KhrVau2012PRE,VauKos2014PLA_1,VauKos2014PLA_2,Rad2012PRE} and 
3D~\cite{AueFre2002JPCM,%
AzhBauRyc2000JCP,HynDij2003JPCM,DavKohMoh2000PRE,DixZuk2003JPCM,%
HeiHolBan2011JCP,ZhoZha2003JPCB,KreRob1986PRL,LinBlaDij2013JCP,ColDij2011JCP,%
GapNagPat2012JCP,GapNagPat2014JCP,KvtKww2014PSSB,PauKah2009JPCM,MeiAzh2007JCP}. 
Recently the elastic properties of three dimensional monodisperse Yukawa system 
were studied and it was shown that the system is \emph{partially
auxetic}~\cite{KvtKww2014PSSB}.

The term \emph{auxetic}~\cite{Eva91ENS} refers to the class of materials and
models that exhibit unusual elastic properties. They expand (shrink)
transversally when stretched (compressed) longitudinally. Their Poisson's
ratio~\cite{LanETAL93BOOK} (in contrast to common materials) is \emph{negative}.
Due to their exceptional properties, auxetic materials and models gather
ever growing interest in recent years~\cite{KWW2014PSSB-preface}.

It is worth noting that usually colloidal systems are not composed of
identical particles. Thus it is meaningful to consider the enhancement of the
model by introduction of the particle size dispersion~\cite{LinBlaDij2013JCP,%
ColDij2011JCP}. The study shows that the dispersion in
sizes of particles can significantly impact the
thermodynamic~\cite{PhaRusChe1996PRE,PhaRusZhuCha98JCP} 
(especially elastic~\cite{KwwJwn2006RAMS})
properties of the system. The study of polydisperse hard disks system not only
indicated a strong dependence of elastic constants on particles size
polydispersity, but also revealed surprising behaviour of the Poisson's ratio
in the close--packing limit~\cite{KVTKWW12JCP}. In this context, the subject of
this work are both mono-- and polydisperse systems.

This work encompasses the results of study performed on two dimensional
hexagonal systems composed of Yukawa particles. The elastic properties of
mono-- and polydisperse two--dimensional crystals of hard--core repulsive
Yukawa particles are described. Beside studies of the influence of
screening length, the temperature and the particles size polydispersity on
elastic properties, the authors posed additional question: whether this
system will also exhibit auxetic properties. It is worth to add that the
elastic properties of such a system have never been investigated.

The structure of this manuscript is the following. In section~\ref{sec_prel}
basic formulas and definitions of physical quantities necessary to describe
the elasticity of the studied system, as well as the description of models
and methods are provided. In section~\ref{sec_res_dis} the simulation results
and their discussion are presented. The last section~(\ref{sec_summ}) provides
the summary of the work.

\begin{figure}[t]%
\includegraphics[width=0.235\textwidth]{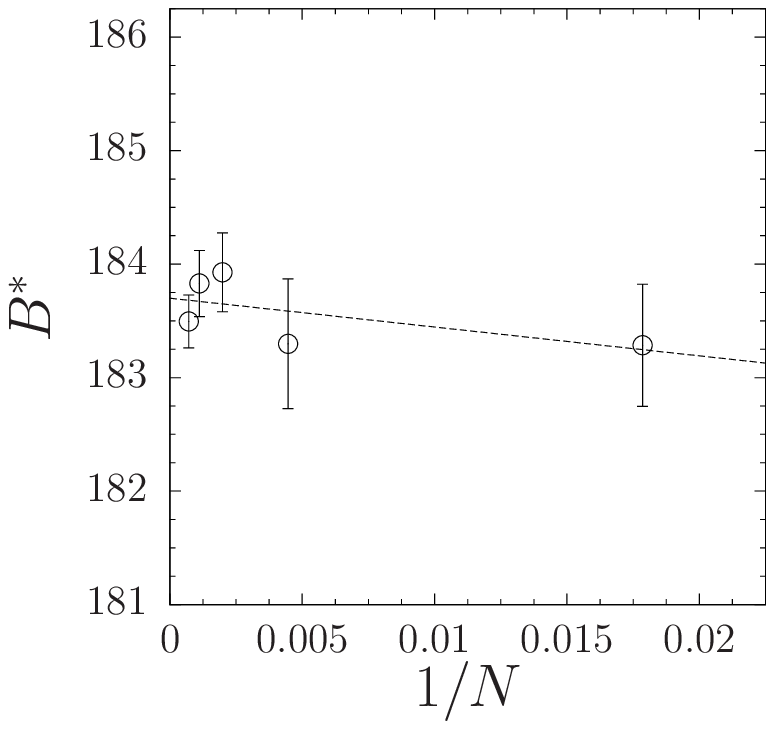}\hfill
\includegraphics[width=0.235\textwidth]{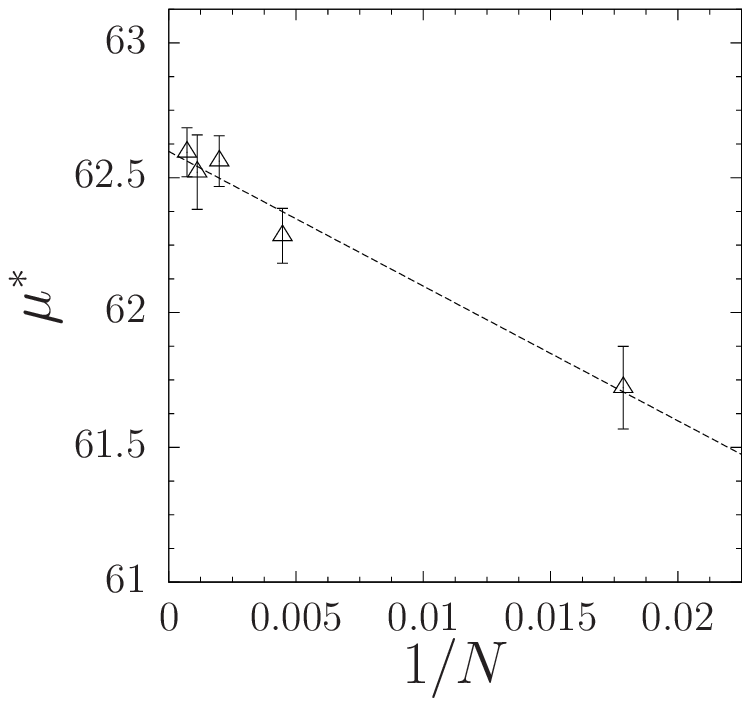}\hfill
\includegraphics[width=0.235\textwidth]{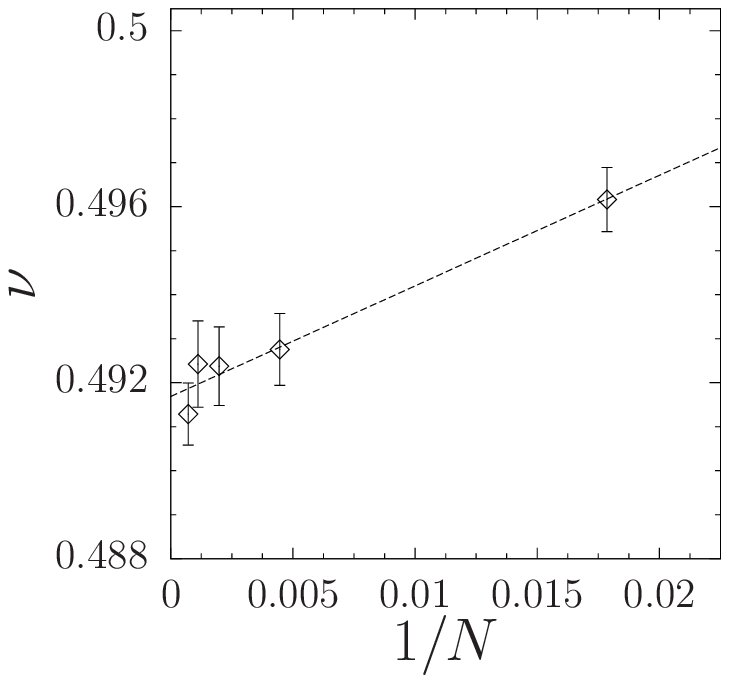}\hfill
\includegraphics[width=0.235\textwidth]{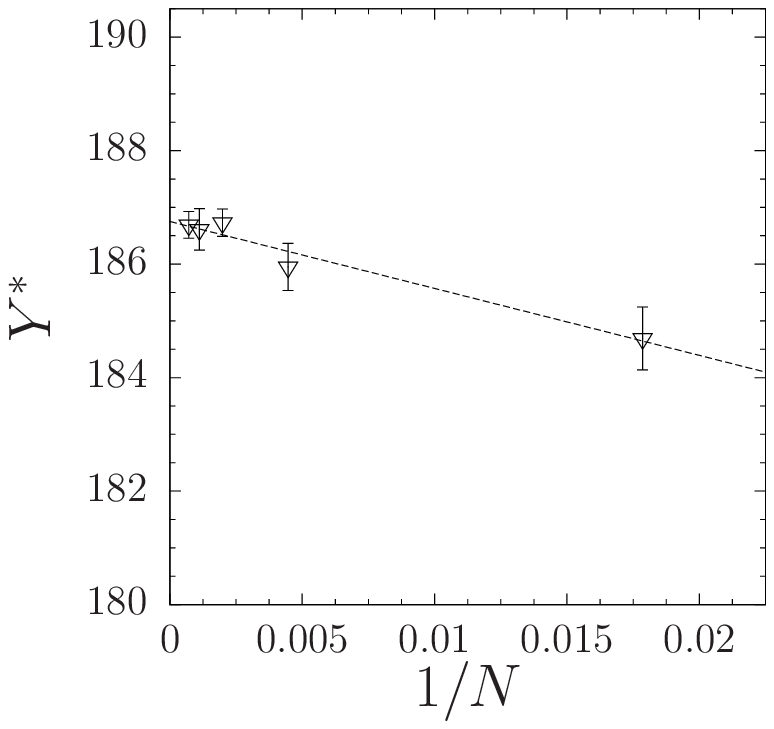}
 \caption{Size dependence of the elastic moduli $B$~(a), $\mu$~(b), Poisson's
 ratio~(c), and Young's modulus~(d) for the system consisting of $N=56,\ 224,
 \ 504,\ 896,\ 1400$ particles, at $p^*=30$, $\beta\epsilon=20$ and 
 $\kappa\sigma=10$. The lines represent linear fits to the obtained data.
 }
 \label{fig1}
\end{figure}
\begin{figure}[t]%
\includegraphics[width=0.235\textwidth]{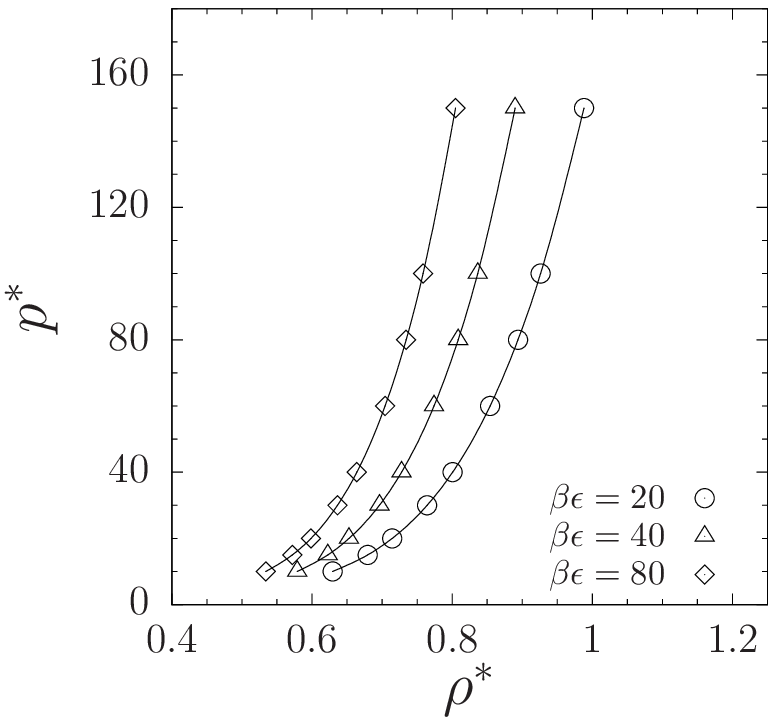}\hfill
\includegraphics[width=0.235\textwidth]{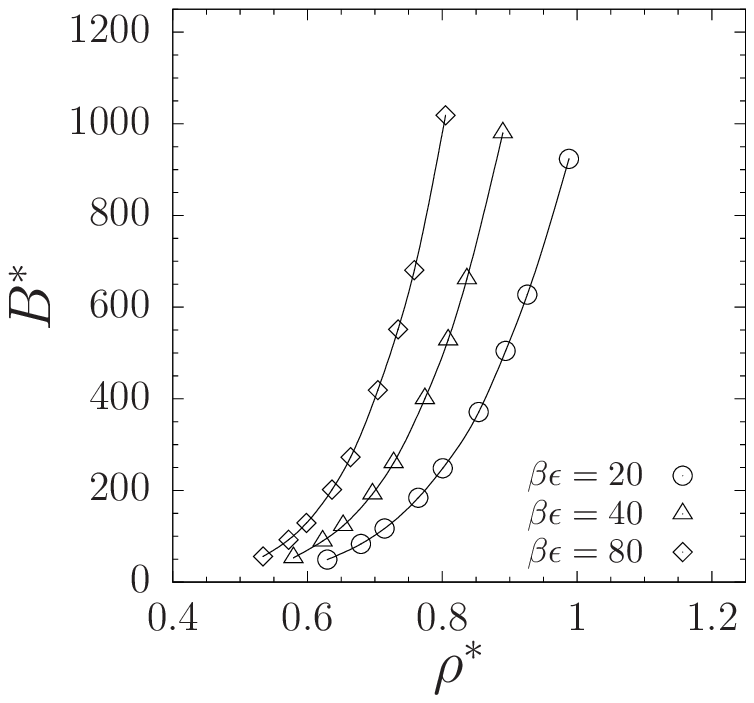}\hfill
\includegraphics[width=0.235\textwidth]{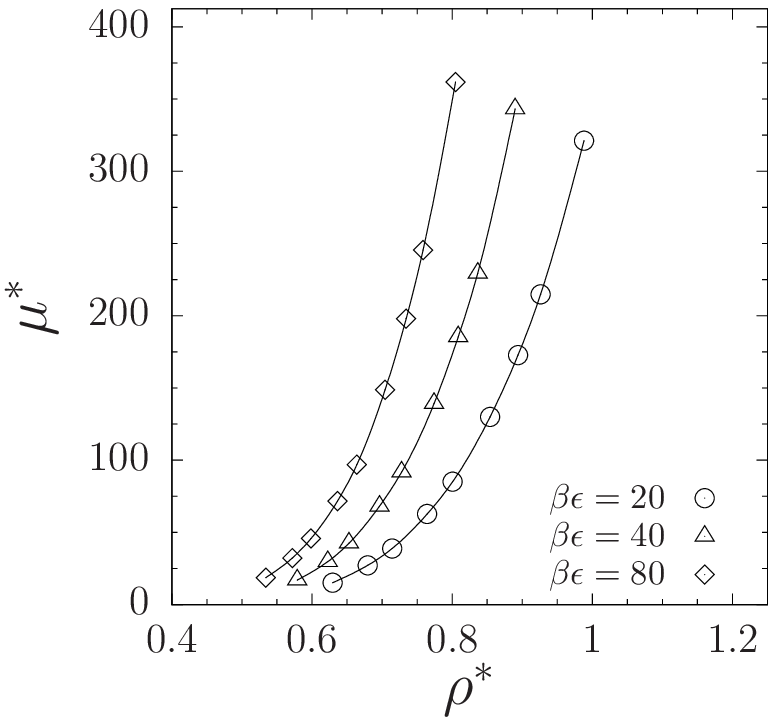}\hfill
\includegraphics[width=0.235\textwidth]{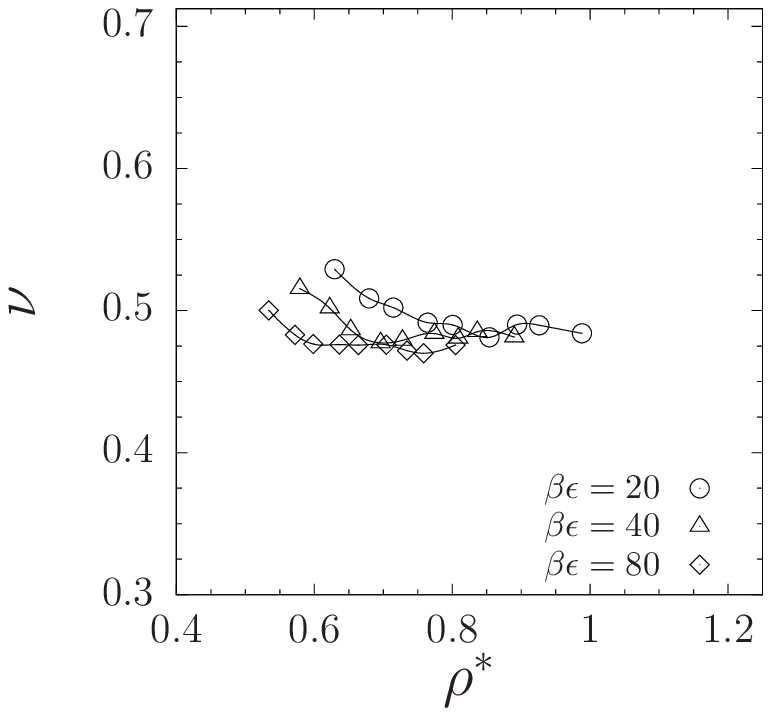}
 \caption{Thermodynamic characteristics of the monodisprese hard--core repulsive
    Yukawa system at $\kappa\sigma=10$. The pressure $p^*$(a); bulk modulus
    $B^*$(b); shear modulus $\mu^*$(c) and the Poisson's ratio $\nu$(d) are
    shown as functions of density for three selected values of temperatures.
    The lines are drawn to guide the eye.
    }
 \label{fig2}
\end{figure}
\begin{figure}[t]
\includegraphics[width=0.235\textwidth]{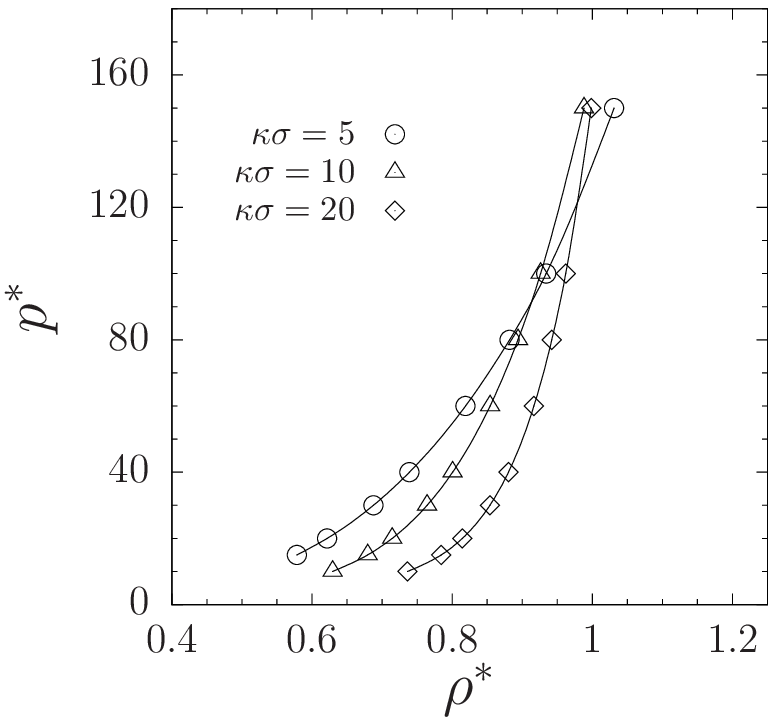}\hfill
\includegraphics[width=0.235\textwidth]{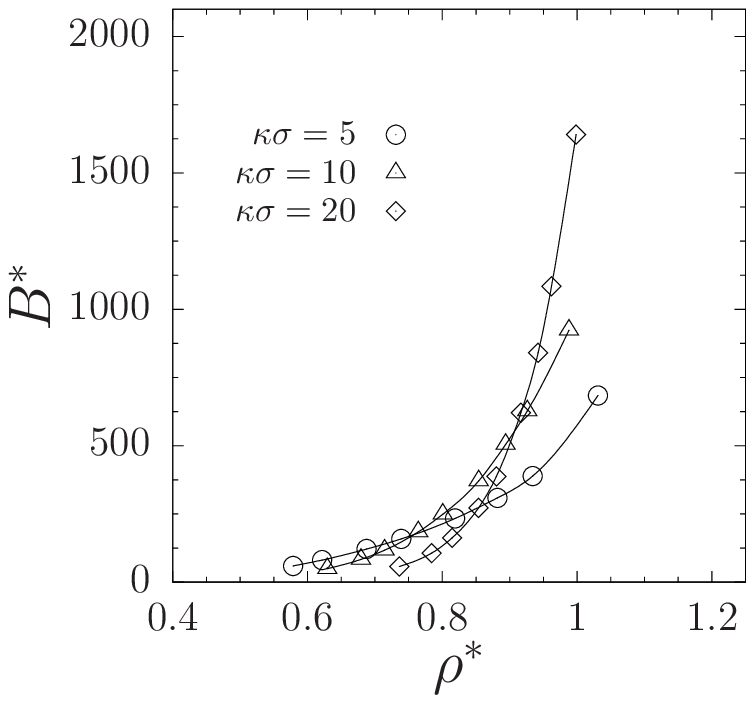}\hfill
\includegraphics[width=0.235\textwidth]{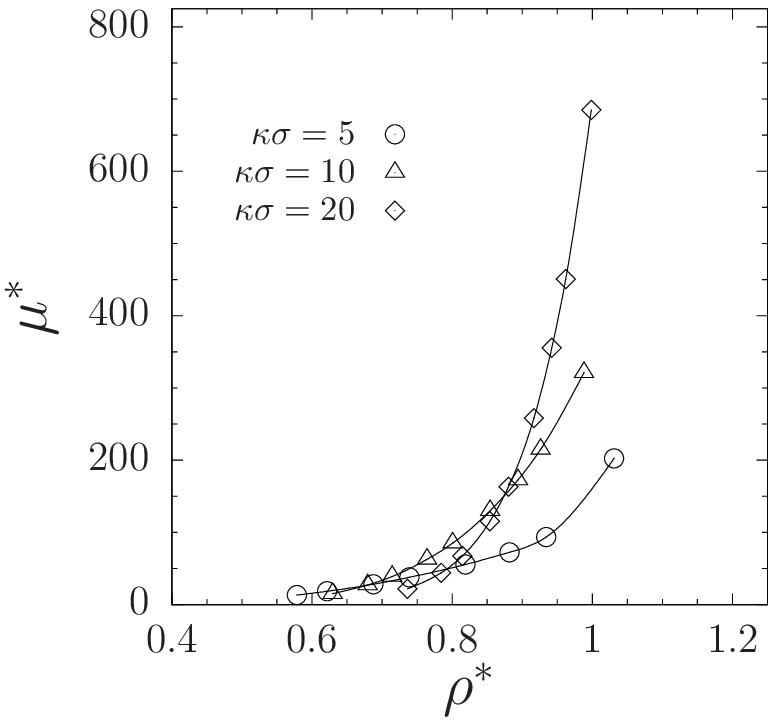}\hfill
\includegraphics[width=0.235\textwidth]{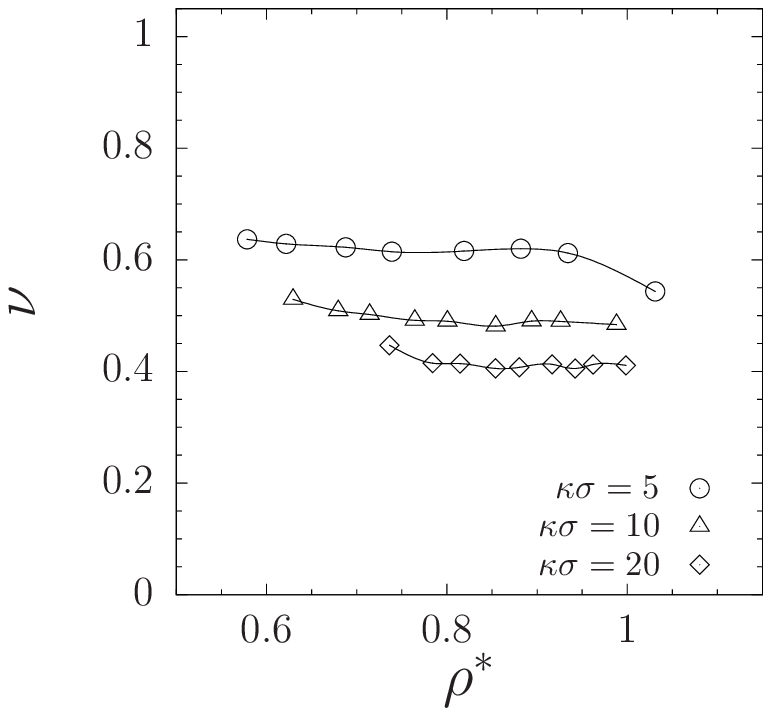}
\caption {Thermodynamic characteristic of the monodisperse hard--core
repulsive Yukawa system at $\beta\epsilon=20$ as functions of density: (a)
pressure $p^*$, (b) the bulk modulus $B^*$, (c) the shear modulus $\mu^*$, and
(d) the Poisson's ratio. The lines are drawn to guide the eye.}
\label{fig3}
\end{figure}

\section{Preliminaries}
\label{sec_prel}

 \subsection{Elastic properties}
 of the system under  non--zero external pressure $p$ which is subjected to
 reversible
 deformation can be described by expanding the free  enthalpy (Gibbs free
 energy) with respect to the Lagrange strain tensor components. The change of
 the Gibbs free energy corresponding the elastic deformation of hexagonal
 lattice under external pressure $p$ expressed with respect to the macroscopic
 elastic moduli, is of the form:
 \begin{eqnarray}
  \Delta G/V_{p}&=&\frac{1}{2}B(\varepsilon_{xx}+ \varepsilon_{yy})^2
  \nonumber\\
  &+& \frac{1}{2}\mu[4\varepsilon_{xy}^2 + (\varepsilon_{xx} -
  \varepsilon_{yy})^2]\ ,
  \label{eq:Gexp2}
 \end{eqnarray}
 where $\varepsilon$ is the strain tensor; $V_{p}$ is the reference volume (at
 equilibrium) of the system under pressure $p$; $B$ is the bulk modulus, and
 $\mu$ is the shear modulus~\cite{KWW89PLA}. It follows from 
 eq.~(\ref{eq:Gexp2}) that hexagonal lattice is elastically 
 isotropic~\cite{LanETAL93BOOK}.

 \subsection{The Poisson's ratio}
 is the negative ratio of the relative changes of transverse to longitudinal
 dimensions when only the longitudinal stress tensor component is infinitesimally
 changed. For hexagonal structure in two dimensions under pressure $p$ one can
 express this quantity in the form~\cite{KWW89PLA}:
 \begin{equation}
  \nu=\frac{B-\mu}{B+\mu} \label{eq:pois}.
 \end{equation}
 It is worth adding that for 2D systems the Poisson's ratio can change in the
 range of $-1\le\nu\le 1$~\cite{KWW89PLA}.

 \subsection{Particle size dispersion} was introduced randomly according to the
 normal distribution with the first two central moments equal to $\sigma$ and
 $\delta$, respectively.
 The standard deviation of the particle size distribution divided by the mean
 diameter of the particle, called the \emph{polydispersity
 parameter}~\cite{PhaRusZhuCha98JCP}:
 \begin{equation}
  \delta=\frac{1}{\left<\sigma\right>}
         \sqrt{\left<\sigma^2\right>-\left<\sigma\right>^2}\ ,
 \end{equation}
 was used to control the degree of disorder in the system.
 It should be noted that, since the system consists of finite
 number of particles, it is important to control the values of
 $\left<\sigma\right>$ and $\delta$ as well as possible.
 The simplest way to meet
 this condition is to repeatedly generate different sets of random numbers
 until the one is found whose properties fall, within a specified accuracy,
 near the desired values~\cite{KvtKww2013PSSB}. In this work, however, a
 different approach was used. After generating a set of $n$ Gaussian pseudo
 random numbers (using method described in~\cite{AllTil1987BOOK}), the
 corresponding average $\overline{\sigma}^{(n)}$ and standard deviation
 $\delta^{(n)}$ are used to convert this set to the \emph{standard normal
 distribution} $Z=(X-\overline{\sigma}^{(n)})/\delta^{(n)}$. Next one can
 convert obtained values such that they exactly match any desired distribution
 $X'=(Z-\sigma)/\delta$, with the machine accuracy. This approach
 ensures the compatibility of distributions for different samples with very
 high accuracy.

 \subsection{The model description}
  starts with introducing reduced thermodynamic quantities used in the simulations:
 \begin{center}
  \begin{tabular}{lll}
   \hline\hline
    reduced value & symbol & therm. equiv. \\
   \hline
    inverse temperature & $\beta\epsilon$ & $\epsilon/k_BT$ \\
    number density &$\rho^*$ & $\rho\sigma^2$\\
    pressure & $p^*$ & $\beta P \sigma^2$\\
    bulk modulus &$B^*$&$B\sigma^2/k_BT$\\
    shear modulus &$\mu^*$&$\mu\sigma^2/k_BT$ \\
   \hline
  \end{tabular}
 \end{center}
 where $T$ is the temperature and $k_B$ is the Boltzman constant.
 The studied system consists of $N$ particles interacting via HCRYP. For the
 monodispersed systems, the potential of interaction is of the form:
 \begin{equation}
  \beta u(r)=
    \begin{cases}
     \infty, & r<\sigma, \\
     \beta\epsilon \frac{\exp[-\kappa (r-\sigma)]}{r/\sigma},& r\geq\sigma,
    \end{cases}
  \label{eq:yuk-mono}
 \end{equation}
 where $r$ is the distance between the centers of particles, $\sigma$ is the
 diameter of the hard core, $\epsilon$ is the value of pair interaction of
 particles at contact (i.e. at $r=\sigma$) and $\kappa^{-1}$ is the
 Debye screening length. In the case of systems with non--zero size
 polydispersity of particles a generalization of the
 equation~(\ref{eq:yuk-mono}) proposed by Colombo and
 Dijkstra~\cite{ColDij2011JCP} was used:
 \begin{equation}
  \beta u(r)=
   \begin{cases}
    \infty, & r<\overline{\sigma}_{ij}, \\
    \beta\epsilon \frac{\sigma_i\sigma_j}{\overline{\sigma}r}
        \exp[-\kappa (r-\overline{\sigma}_{ij})],& r\geq\overline{\sigma}_{ij},
   \end{cases}
  \label{eq:yuk-poly}
 \end{equation}
 
  \subsection{Computer simulations} have been performed in the
 isothermal--isobaric ensemble ($NpT$) by Monte Carlo method. In order to
 calculate the elastic moduli, the approach based on the averaging
 of strain fluctuations proposed by Parrinello--Rahman~\cite{ParRah82JCP} and
 further developed by Ray and Rahman~\cite{RayRah84JCP,RayRah85JCP}, was used.
 The details of the method were described in ref.~\cite{KwwKvtMko2003PhysRevE}. 
 It is worth noting that in the
 case of a potential with long--range interactions the minimum image method
 (MIM)~\cite{KvtKww2014CPC} can be used. Our studies showed that to obtain
 accurate results for following parameters ($\kappa\le 4,\ \beta \ge 80$,
 $r_\textrm{cut}\le2.5$) of studied potential the MIM or another method, which
 takes into account the long--range interactions, must be
 used.

 The typical
 studied systems were composed of $N=224$ particles under periodic boundary
 conditions. Samples for simulation were chosen such that the edge ratio was
 as close to $1$ as possible. The particles were placed in the $14a \times
 8\sqrt{3}a$ box, where $a$ is the distance between the nearest neighbors in
 crystal. The term \emph{nearest neighbor} refers to particles that share
 a common edge of their Dirichlet polygons. Results from simulations were
 averaged over $10^7$ Monte Carlo cycles after prior equilibration of
 $10^6$ cycles. To check the dependence of the elastic moduli on the size of
 studied systems, the system of $N=56,\ 224,\ 504,\ 896,\ 1400$ particles have
 been studied, for which $5\cdot 10^6,\ 10^7,\ 3\cdot10^7,\ 6\cdot10^7,\ 10^8$
 MC cycles were performed.
 The interaction potential implemented for simulations of
 polydisperse systems is described by the equation~(\ref{eq:yuk-poly}). For
 monodisperse systems the latter reduces to the form of eq.~(\ref{eq:yuk-mono}).
 Elastic moduli for each phase point were determined by averaging over $10$
 independent simulations, each of which was performed for a different structure
 generated from a proper distribution.

\section{Results and discussion}
\label{sec_res_dis}

\begin{figure}[htpb]
\includegraphics[width=0.300\textwidth]{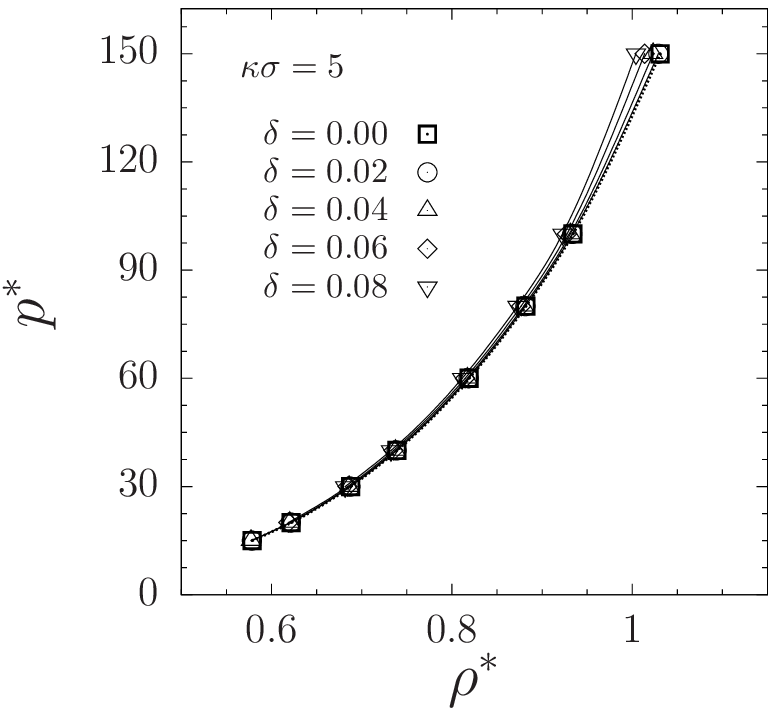}\hfill
\includegraphics[width=0.300\textwidth]{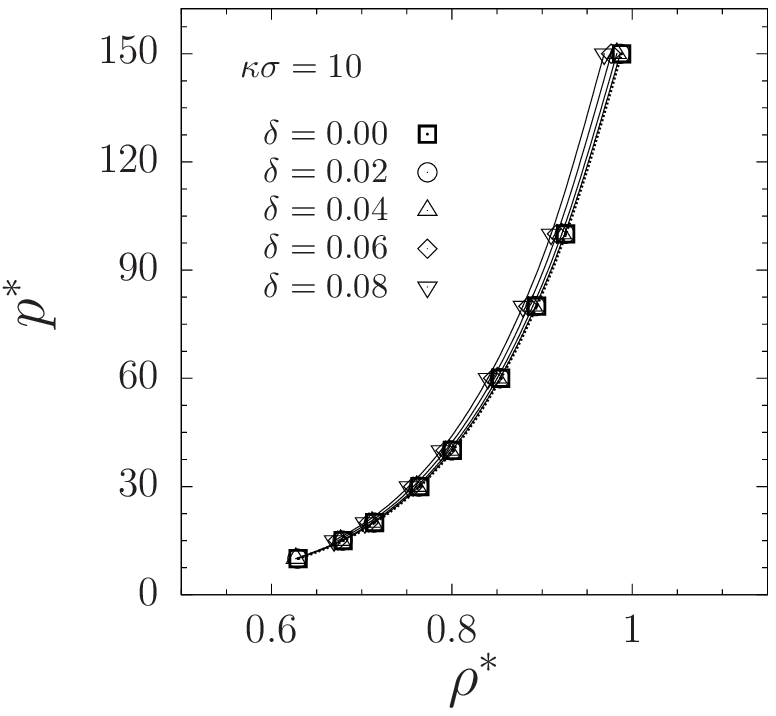}\hfill
\includegraphics[width=0.300\textwidth]{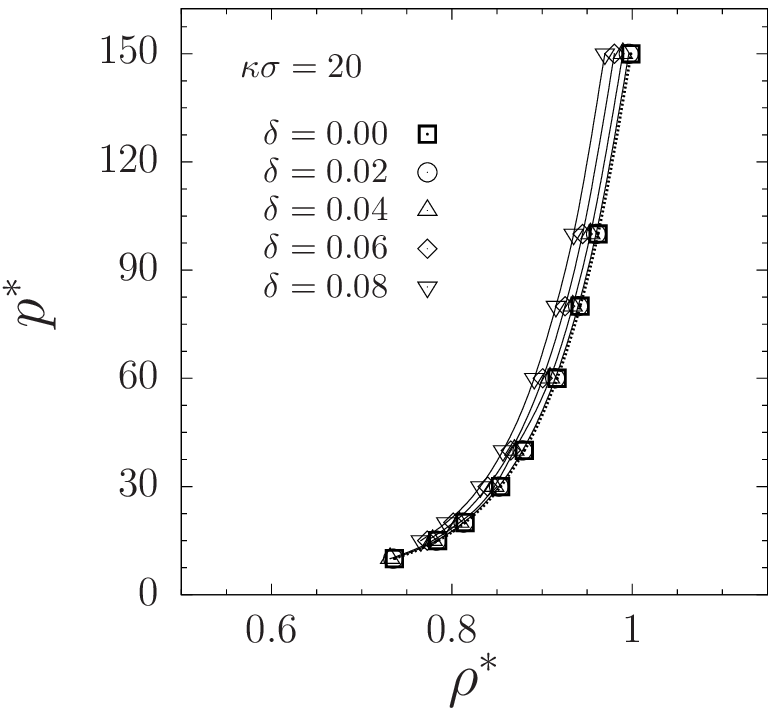}
\caption{Equation of state of the hard--core repulsive Yukawa system at
$\beta\epsilon=20$ for  $\kappa\sigma=5$~(a), $\kappa\sigma=10$~(b), and 
$\kappa\sigma=20$~(c). The lines are drawn to guide the eye.}
\label{fig4}
\end{figure}
\begin{figure}[htpb]
\includegraphics[width=0.300\textwidth]{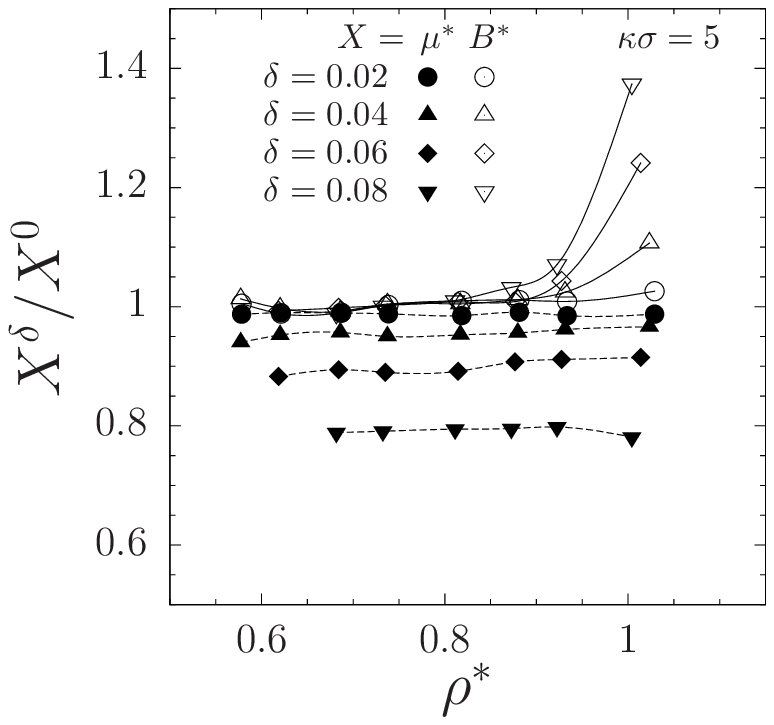}\hfill
\includegraphics[width=0.300\textwidth]{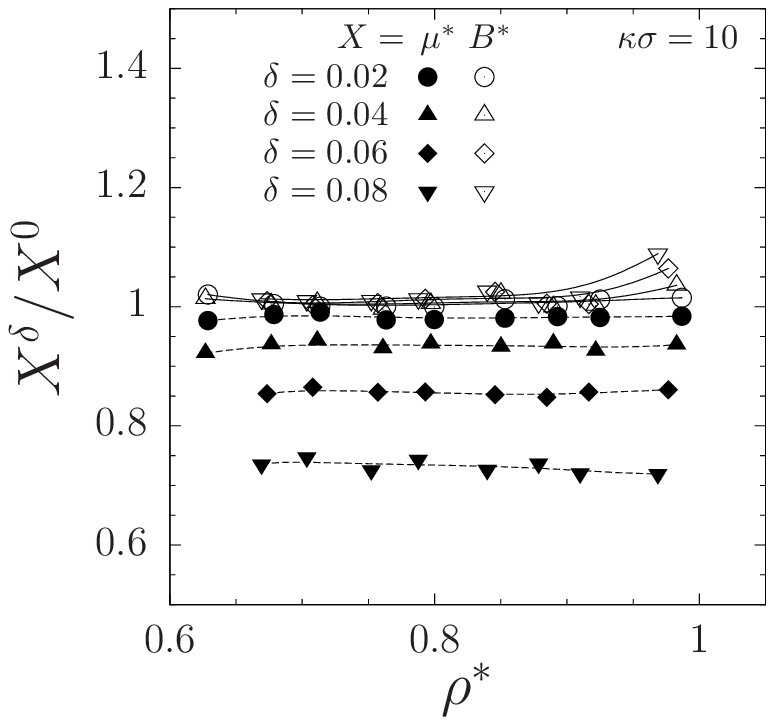}\hfill
\includegraphics[width=0.300\textwidth]{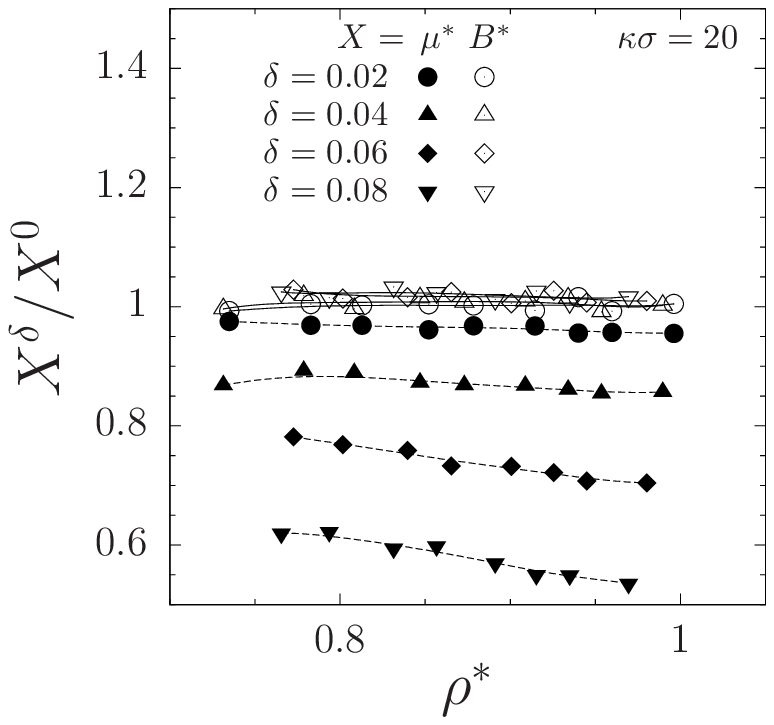}
\caption {Relative change of elastic moduli due to the non--zero size
polydispersity for
$\kappa\sigma=5$~(a), $\kappa\sigma=10$~(b) and $\kappa\sigma=20$~(c).
$X$ stands for $B$ or $\mu$. The lines are drawn to guide the eye.}
\label{fig5}
\end{figure}
\begin{figure}[htpb]
\centering
\includegraphics[width=0.300\textwidth]{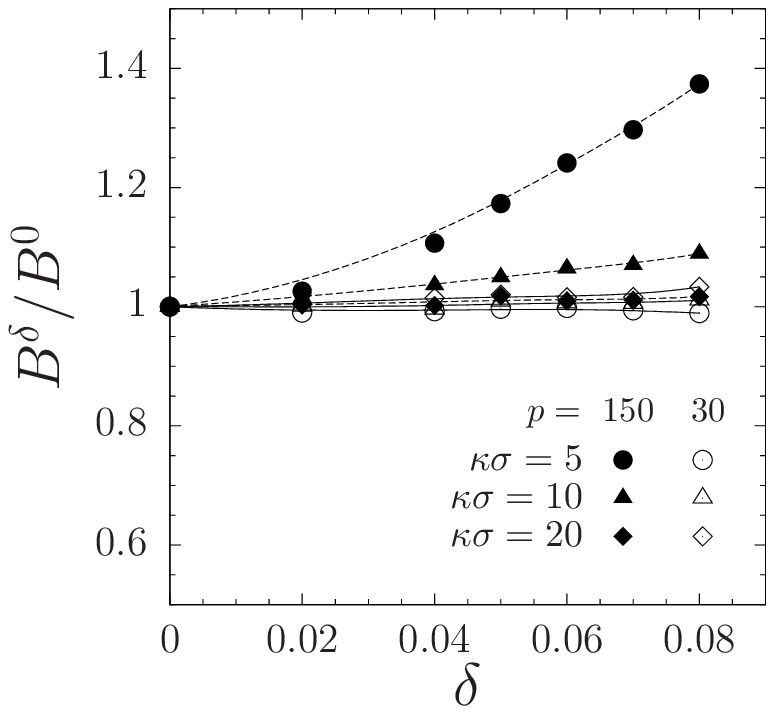}
\includegraphics[width=0.300\textwidth]{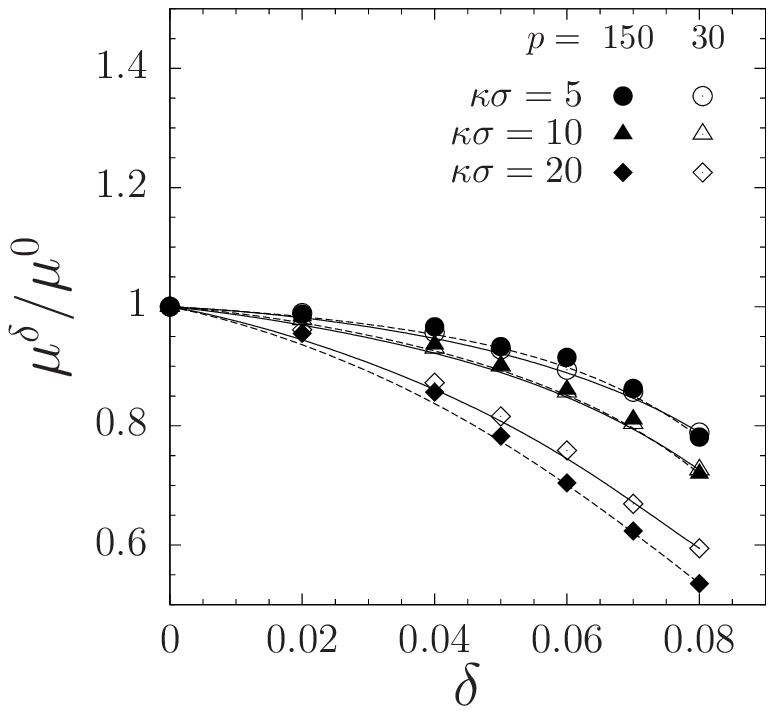}
\caption {
Relative change of (a) the bulk modulus and (b) the shear modulus
due to the non--zero size polydispersity, studied at $\beta\epsilon=20$,
for two different values of pressure. The lines are drawn to guide the eye.
}
\label{fig6}
\end{figure}

Based on the equation of state in Ref.~\cite{AsgDavTan2001PRE}, values of the 
potential's parameters corresponding to solid phases were chosen to allow for
comparison of simulation results with experimental data, e.g. $\beta \varepsilon 
\approx 20$ corresponds to colloidal systems in a low dielectric solvent at 
room temperature~\cite{ColDij2011JCP}. The discussion of 
elastic properties of given systems that follows, is presented in the form of 
the three cases: the constant screening length $(\kappa\sigma)^{-1}$ 
(sec.~\ref{subsec_KS}); the constant temperature $(\beta\epsilon)^{-1}$ 
(sec.~\ref{subsec_BE}); and the influence of particle size polydispersity 
(sec.~\ref{subsec_P}) on the mentioned above properties.

In Figure~\ref{fig1} are shown dependencies of the elastic moduli and
Poisson's ratio with respect to the size of the system. It may be noted that the
simulations performed for $N=224$ yield results that differ only a few percent
from the values extrapolated to the thermodynamic limit ($N\rightarrow\infty$). 
One can add that the agreement is quite good for all the studied system sizes. 
In view of the above, all further discussion applies to systems containing 
$N=224$ particles.

 \subsection{The case of constant screening length}
 \label{subsec_KS}
 is illustrated in Figure~\ref{fig2}. The elastic moduli as functions of 
 density for  $(\kappa\sigma)^{-1}=10^{-1}$ are presented there. 
 One can observe a monotonic increase of both elastic moduli and pressure with 
 increasing density. A decrease of the temperature results in an increase of 
 pressure and elastic moduli at entire range of studied densities. Furthermore, 
 it causes a shift of the entire isotherm describing the lower temperature in 
 the direction of high densities.

 The density dependence of Poisson's ratio is presented in Fig.~\ref{fig2}d.
 It is immediately apparent that for selected $\kappa$ values of Poisson's
 ratio for different temperatures are very close one to another at high
 densities. This suggests that the influence of particles' hard cores becomes 
 dominant at these densities. At low densities an increase of temperature causes 
 an increase of the Poisson's ratio, however the dependence on 
 temperature is rather weak.
\begin{figure}[htpb]
\includegraphics[width=0.300\textwidth]{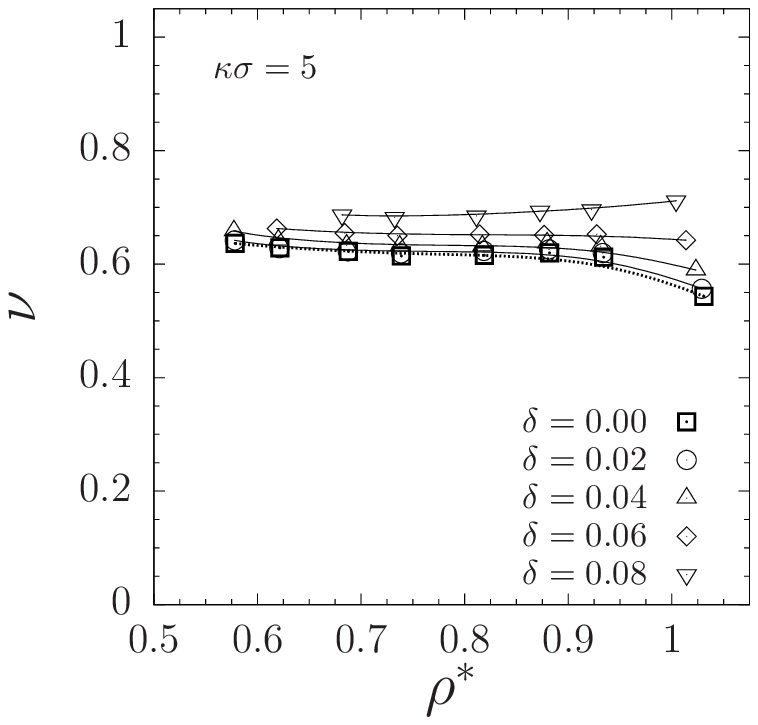}\hfill
\includegraphics[width=0.300\textwidth]{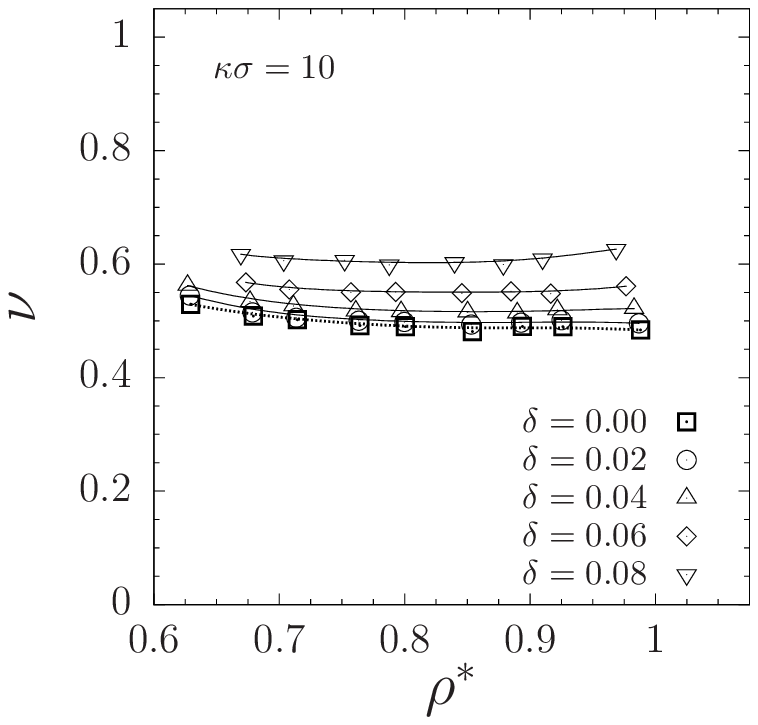}\hfill
\includegraphics[width=0.300\textwidth]{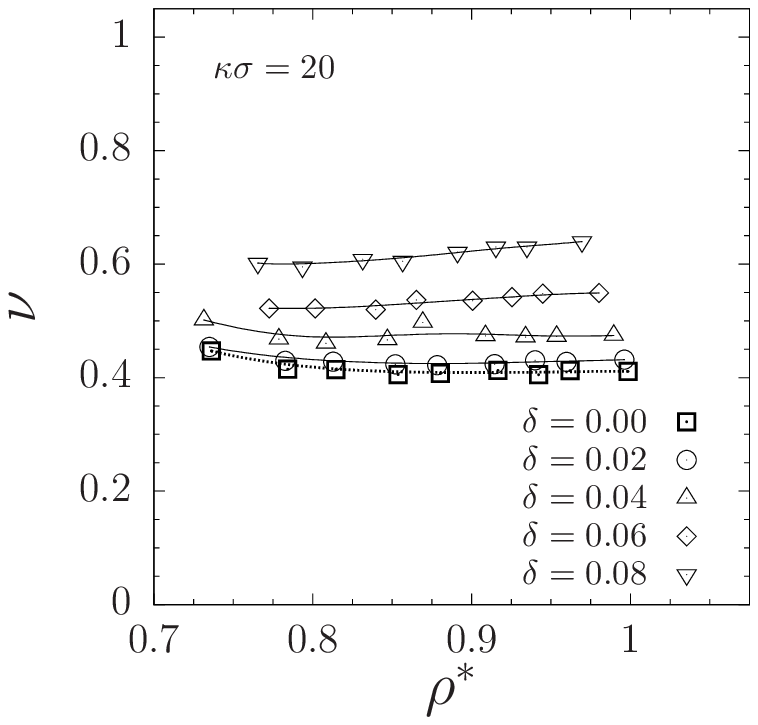}
\caption {Density dependence of the Poisson's ratio of the polydisperse 
hard--core repulsive Yukawa system at $\beta\epsilon=20$,
for three values of the screening length $\kappa\sigma=5$~(a), 
$\kappa\sigma=10$~(b), and  $\kappa\sigma=20$~(c).}
\label{fig7}
\end{figure}

 \subsection{The case of constant temperature}
 \label{subsec_BE}
 was investigated for $(\beta\epsilon)^{-1}=20^{-1}$.
 Figure~\ref{fig3} illustrates the density dependence of pressure~(a) as 
 well as the bulk modulus~(b), the shear modulus~(c) and the Poisson's 
 ratio~(d).  
 Similarly to previous case, an increase of the pressure and the elastic moduli 
 with the density can be observed. However, the growth rate
 of the elastic moduli is different and it clearly depends on the screening length.
 An increase of the elastic moduli with increasing density is slower for 
 long--range 
 potential ($(\kappa\sigma)^{-1}=5^{-1}$) than for interactions with shorter 
 screening length ($(\kappa\sigma)^{-1}=20^{-1}$).

 Figure~\ref{fig3}d illustrates the dependence of Poisson's ratio on density for
 systems with different screening lengths. It is noticeable that for a given
 $\kappa$ the Poisson's ratio weakly depends on the density of the system.
 However, the screening length itself essentially influences the value of
 Poisson's ratio. This leads to conclusion that one can control the Poisson's
 ratio by modifying the Debye screening length (e.g. by changing the properties
 of the electrolyte).

 \subsection{Influence of size polydispersity}
 \label{subsec_P}
 of the particles' hard cores on elastic properties is discussed in this
 section. One can start from the examination of the equation of state presented
 in the Figure~\ref{fig4}. As in the previous case, presented data correspond
 to systems at reduced temperature $(\beta\epsilon)^{-1}=20^{-1}$ for three
 values of screening length $(\kappa\sigma)^{-1}$: $5^{-1}, 10^{-1}$, and
 $20^{-1}$. 
 The choice of these values is dictated by two factors. Firstly, studies of 
 Poisson's ratio in case of monodisperse system (see sec.~\ref{subsec_KS}) 
 showed that its temperature dependence is rather week while the screening 
 length has a significant impact on its value. Secondly, the present study is 
 mainly focused on looking for mechanisms affecting Poisson's ratio. Thus the 
 influence of the screening length on the elastic properties and Poisson's
 ratio in systems with non-zero polydispersity of particles sizes was considered 
 at a selected (constant) temperature. Systems with four 
 different values of polydispersity parameter $\delta$ have been investigated. 
 Its influence manifests as the change of number density
 (i.e. the ratio of the number of particles per 2D 'volume') for a given 
 pressure (see Fig.~\ref{fig4}). An increase of the size polydispersity 
 causes a decrease of the number density at a given pressure. The decrease, 
 however, depends on the screening length. For long--range interactions 
 (Fig.~\ref{fig4}a) the impact of size dispersion is significantly smaller and
 noticeable only at high pressures. Once the screening length is getting smaller
 (increasing $\kappa$), rising $\delta$ enforces the system to expand even at
 lower pressures.

 Figure~\ref{fig5} illustrates changes of elastic moduli caused by the size
 polydispersity relative to monodisperse system for different screening lengths.
 The relations are presented as functions of density. One can notice that
 the influence of the size polydispersity is very different for both the 
 moduli. In the case of the bulk modulus, a significant increase can be observed
 only for high densities and long interactions (Fig~\ref{fig5}a). For lower
 densities the influence of the size polydispersity on the bulk modulus is
 negligible. Furthermore, when the effect of particles' hard cores becomes 
 dominant (increasing $\kappa$), one can observe only minor impact of the 
 non--zero polydispersity on $B$ in the whole range of studied densities 
 (Fig.~\ref{fig5}c). This effect is further illustrated in the 
 Figure~\ref{fig6}a, where the relative bulk modulus with respect the 
 polydispersity parameter is presented for three screening lengths at high and 
 low pressures. It can be seen there that only for long--range interactions and 
 high pressure one can observe a significant influence of polydispersity on the 
 bulk modulus, and this impact quickly decays with the decrease of the screening 
 length.

 The effect of the size polydispersity on the shear modulus $\mu$ is stronger 
 than on  the bulk modulus. In Figure~\ref{fig5}a one can see that an increase 
 of $\delta$ decreases the shear modulus. From the other hand shortening of the
 screening length increases an influence of the polydispersity on the shear
 modulus~(Fig.~\ref{fig5} and~\ref{fig6}). Thus, one can conclude that for a 
 system with given size polydispersity of particles one can additionally control 
 its (the system) susceptibility to the shear deformations by changing the Debye 
 screening length (e.g. by modifying the properties of the electrolyte). Unlike 
 the bulk modulus, pressure has a little impact on a decrease of $\mu$ 
 with the increasing parameter of polydispersity (Fig.~\ref{fig6}b). Only in the
 case of shortest screening lengths ($\kappa\sigma=20$) the difference between 
 the considered properties of systems under low and high pressures is 
 noticeable. For smaller values of $\kappa$ only an increase of the 
 polydispersity parameter is responsible for a decrease of the shear modulus. It
 is worth noting, that  most significant decrease of the shear modulus is 
 observed for the shortest screening length.

 Figure~\ref{fig7} shows the density dependence of the Poisson's ratio
 for different
 values of polydispersity parameter and various screening lengths. It can be seen
 that for long-range interactions the size polydispersity of particles affects
 weakly the Poisson's ratio and it weakly depends on the density. With
 shortening the screening length, the impact of polydispersity becomes more
 noticeable, getting very significant role for the short-range interactions.

\section{Summary and conclusions}
\label{sec_summ}
The influence of the temperature, screening lengths and size polydispersity on
elastic properties of the two-dimensional hexagonal structure of hard--core, 
repulsive
Yukawa particles was investigated by Monte Carlo simulations in the $NpT$
ensemble with variable shape of the periodic box. The obtained results show that 
influence of the screening length on the elastic properties of the studied 
systems is significant. It was also shown that size polydispersity impacts the 
elastic properties, especially the shear modulus, which decreases with 
decreasing the size polydispersity parameter. However, an increase of particle 
size dispersion is reflected in the increase of Poisson's ratio. It was shown 
that the 2D Yukawa model is non--auxetic but the value of the Poisson's ratio 
decreases with decreasing screening length. This mechanism can be useful in 
designing systems with low Poisson's ratio.

\section*{Acknowledgement}
We are grateful to Dr. Miko{\l}aj Kowalik (Pennsylvania State University) for 
discussions. This work was supported partially by the Polish National Science
Center grants
DEC-2012/05/B/ST3/03255. The calculations were partially performed at Pozna\'n
Supercomputing and Networking Center (PCSS).

\bibliographystyle{plain}


\begin{thebibliography}{[10]}

\bibitem{GasRus98PT}
 \textsc{A.\,P. Gast} and  \textsc{W.\,B. Russel},
 \jr{Phys. Today} \textbf{51}, 24 (1998).


\bibitem{RusETAL91BOOK}
 \textsc{W.\,B. Russel},  \textsc{D.\,A. Saville},  and  \textsc{W.\,R.
  Schowalter},
Colloidal Dispersions (Cambridge University Press, Cambridge, 1991).


\bibitem{Pie80PRL}
 \textsc{P.~Pieranski},
 \jr{Phys.\ Rev.\ Lett.} \textbf{45}, 569 (1980).


\bibitem{AueFre2002JPCM}
 \textsc{S.~Auer} and  \textsc{D.~Frenkel},
 \jr{J. Phys.: Condens. Matter} \textbf{14}, 7667--7680 (2002).


\bibitem{AzhBauRyc2000JCP}
 \textsc{F.\,E. Azhar},  \textsc{M.~Baus},  and  \textsc{J.\,P. Ryckaert},
 \jr{J. Chem. Phys} \textbf{112}, 5121--5126 (2000).


\bibitem{HynDij2003JPCM}
 \textsc{A.\,P. Hynninen} and  \textsc{M.~Dijkstra},
 \jr{J. Phys.: Condens. Matter} \textbf{15}, 3557--3567 (2003).


\bibitem{AsgDavTan2001PRE}
 \textsc{R.~Asgari},  \textsc{B.~Davoudi},  and  \textsc{B.~Tanatar},
 \jr{Phys. Rev. E} \textbf{64}, 41406 (2001).


\bibitem{HynDij03PRE}
 \textsc{A.\,P. Hynninen} and  \textsc{M.~Dijkstra},
 \jr{Phys. Rev. E} \textbf{68}, 021407 (2003).


\bibitem{AssMesLow2008JCP}
 \textsc{L.~Assoud},  \textsc{R.~Messina},  and  \textsc{H.~Lowen},
 \jr{J. Chem. Phys.} \textbf{129}, 164511 (2008).


\bibitem{DavKohMoh2000PRE}
 \textsc{B.~Davoudi},  \textsc{M.~Kohandel},  \textsc{M.~Mohammadi},  and
  \textsc{B.~Tanatar},
 \jr{Phys. Rev. E} \textbf{62}, 6977 (2000).


\bibitem{DixZuk2003JPCM}
 \textsc{N.\,M. Dixit} and  \textsc{C.\,F. Zukoski},
 \jr{J. Phys.: Condens. Matter} \textbf{15}, 1531--1552 (2003).


\bibitem{GlaLow2012JPCM}
 \textsc{T.~Glanz} and  \textsc{H.~Lowen},
 \jr{J. Phys.: Condens. Matter} \textbf{24}, 464114 (2012).


\bibitem{HarKalDon2005PRE}
 \textsc{P.~Hartmann},  \textsc{G.\,J. Kalman},  \textsc{Z.~Donko},  and
  \textsc{K.~Kutasi},
 \jr{Phys. Rev. E} \textbf{72}, 26409 (2005).


\bibitem{KhrVau2012PRE}
 \textsc{Y.\,V. Khrustalyov} and  \textsc{O.\,S. Vaulina},
 \jr{Phys. Rev. E} \textbf{85}, 46405 (2012).


\bibitem{VauKos2014PLA_1}
 \textsc{O.\,S. Vaulina} and  \textsc{X.\,G. Koss},
 \jr{Phys. Lett. A} \textbf{378}, 719--722 (2014).


\bibitem{VauKos2014PLA_2}
 \textsc{O.\,S. Vaulina} and  \textsc{X.\,G. Koss},
 \jr{Phys. Lett. A} \textbf{378}, 3475--3479 (2014).


\bibitem{Rad2012PRE}
 \textsc{A.~Radzvilavicius},
 \jr{Phys. Rev. E} \textbf{85}, 51111 (2012).


\bibitem{HeiHolBan2011JCP}
 \textsc{M.~Heinen},  \textsc{P.~Holmqvist},  \textsc{A.\,J. Banchio},  and
  \textsc{G.~Nagele},
 \jr{J. Chem. Phys} \textbf{134}, 44532 (2011).


\bibitem{ZhoZha2003JPCB}
 \textsc{S.~Zhou} and  \textsc{X.~Zhang},
 \jr{J. Phys. Chem. B} \textbf{107}, 5294--5299 (2003).


\bibitem{KreRob1986PRL}
 \textsc{K.~Krerner},  \textsc{M.\,O. Robbins},  and  \textsc{G.\,S. Grest},
 \jr{Phys. Rev. Lett.} \textbf{57}, 2694--2697 (1986).


\bibitem{LinBlaDij2013JCP}
 \textsc{M.\,N. van\,der Linden},  \textsc{A.~van Blaaderen},  and
  \textsc{M.~Dijkstra},
 \jr{J. Chem. Phys.} \textbf{138}, 114903 (2013).


\bibitem{ColDij2011JCP}
 \textsc{J.~Colombo} and  \textsc{M.~Dijkstra},
 \jr{J. Chem. Phys} \textbf{134}, 154504 (2011).


\bibitem{GapNagPat2012JCP}
 \textsc{J.~Gapinski},  \textsc{G.~Nagele},  and  \textsc{A.~Patkowski},
 \jr{J. Chem. Phys} \textbf{136}, 24507 (2012).


\bibitem{GapNagPat2014JCP}
 \textsc{J.~Gapinski},  \textsc{G.~Nagele},  and  \textsc{A.~Patkowski},
 \jr{J. Chem. Phys} \textbf{141}, 124505 (2014).


\bibitem{KvtKww2014PSSB}
 \textsc{K.\,V. Tretiakov} and  \textsc{K.\,W. Wojciechowski},
 \jr{Phys. Status Solidi B} \textbf{251}, 383--387 (2014).


\bibitem{PauKah2009JPCM}
 \textsc{G.\,J. Pauschenwein} and  \textsc{G.~Kahl},
 \jr{J. Phys.: Condens. Matter} \textbf{21}, 474202 (2009).


\bibitem{MeiAzh2007JCP}
 \textsc{E.\,J. Meijer} and  \textsc{F.\,E. Azhar},
 \jr{J. Chem. Phys} \textbf{106}, 4678 (2007).


\bibitem{Eva91ENS}
 \textsc{K.\,E. Evans},
 \jr{Endeavour, New Series} \textbf{15}, 170 (1991).


\bibitem{LanETAL93BOOK}
 \textsc{L.\,D. Landau} and  \textsc{E.\,M. Lifshits},
Theory of Elasticity, 3rd Edition (Pergamon Press, Oxford, 1993).


\bibitem{KWW2014PSSB-preface}
 \textsc{K.\,L. Alderson1},  \textsc{A.~Alderson},  \textsc{J.\,N. Grima},  and
   \textsc{K.\,W. Wojciechowski},
 \jr{Phys. Status Solidi B} \textbf{251}, 263--266 (2014).


\bibitem{PhaRusChe1996PRE}
 \textsc{S.\,E. Phan},  \textsc{W.\,B. Russel},  \textsc{Z.~Cheng},
  \textsc{J.~Zhu},  \textsc{P.\,M. Chaikin},  \textsc{J.\,H. Dunsmuir},  and
  \textsc{R.\,H. Ottewill},
 \jr{Phys. Rev. E} \textbf{54}, 6633--6645 (1996).


\bibitem{PhaRusZhuCha98JCP}
 \textsc{S.\,E. Phan},  \textsc{W.\,B. Russel},  \textsc{J.\,X. Zhu},  and
  \textsc{P.\,M. Chaikin},
 \jr{J.\ Chem.\ Phys.} \textbf{108}, 9789 (1998).


\bibitem{KwwJwn2006RAMS}
 \textsc{K.\,W. Wojciechowski} and  \textsc{J.~Narojczyk},
 \jr{Rev. Adv. Mater. Sci.} \textbf{12(2)}, 120--126 (2006).


\bibitem{KVTKWW12JCP}
 \textsc{K.\,V. Tretiakov} and  \textsc{K.\,W. Wojciechowski},
 \jr{J. Chem. Phys.} \textbf{136}, 204506 (2012).


\bibitem{KWW89PLA}
 \textsc{K.\,W. Wojciechowski},
 \jr{Phys.\ Lett.\ A} \textbf{137}, 60--64 (1989).


\bibitem{KvtKww2013PSSB}
 \textsc{K.\,V. Tretiakov} and  \textsc{K.\,W. Wojciechowski},
 \jr{Phys. Status Solidi B} \textbf{250(10)}, 2020--2029 (2013).


\bibitem{AllTil1987BOOK}
 \textsc{M.\,P. Allen} and  \textsc{D.\,J. Tildesley},
Computer Simulation of Liquids (Clarendon Press, Oxford, 1987).


\bibitem{ParRah82JCP}
 \textsc{M.~Parrinello} and  \textsc{A.~Rahman},
 \jr{J.\ Chem.\ Phys.} \textbf{76}, 2662 (1982).


\bibitem{RayRah84JCP}
 \textsc{J.\,R. Ray} and  \textsc{A.~Rahman},
 \jr{J.\ Chem.\ Phys.} \textbf{80}, 4423 (1984).


\bibitem{RayRah85JCP}
 \textsc{J.\,R. Ray} and  \textsc{A.~Rahman},
 \jr{J.\ Chem.\ Phys.} \textbf{82}, 4243 (1985).


\bibitem{KwwKvtMko2003PhysRevE}
 \textsc{K.\,W. Wojciechowski},  \textsc{K.\,V. Tretiakov},  and
  \textsc{M.~Kowalik},
 \jr{Phys. Rev. E} \textbf{67}, 036121 (2003).


\bibitem{KvtKww2014CPC}
 \textsc{K.\,V. Tretiakov} and  \textsc{K.\,W. Wojciechowski},
 \jr{Comput. Phys. Commun.} \textbf{189}, 77--83 (2015).


\end{thebibliography}

\providecommand{\jr}[1]{#1}
\providecommand{\etal}{~et~al.}

\end{document}